\documentclass{ws-procs975x65}

\begin{document}
\title{The origin of the Fermi-LAT $\gamma$-ray background}

\author{Mattia Di Mauro, on behalf of the Fermi-LAT Collaboration}

\address{Kavli Institute for Particle Astrophysics and Cosmology, Department of Physics and SLAC National Accelerator Laboratory, Stanford University, USA. \\
E-mail: mdimauro@slac.stanford.edu}

\begin{abstract}
The origin of the Isotropic Diffuse $\gamma$-Ray Background (IGRB) is one of the most intriguing mystery in astrophysics. Recently the Large Area Telescope (LAT) on board the {\it Fermi} Gamma-Ray Space Telescope has measured the IGRB spectrum from 100 MeV to 820 GeV. Thousands of extragalactic sources are detected at $\gamma$-ray energies and, thanks to {\it Fermi}-LAT catalogs and to population studies, we show that the IGRB can be explained, in the entire energy range, by the $\gamma$-ray emission from unresolved misaligned Active Galactic Nuclei, blazars and Star Forming Galaxies. Finally we derive, with a dedicated analysis based on the 2FHL {\it Fermi}-LAT catalog at $E>50$ GeV, the intrinsic source count distribution of blazars and demonstrate that this source population explains the $86^{+16}_{-14}\%$ of the total extragalactic $\gamma$-ray background.
\end{abstract}


\bodymatter


\section{Introduction}

The $\gamma$-ray sky has been observed since 1972 by the OSO-3 satellite \refcite{1972ApJ...177..341K}. This experiment provided the first $\gamma$-ray sky map with 621 events detected above 50 MeV and it first revealed the existence of an isotropic emission. The presence of this isotropic component, called Isotropic Diffuse $\gamma$-Ray Background (IGRB), then has been measured by SAS-2 \refcite{1975ApJ...198..163F} and the EGRET \refcite{1998ApJ...494..523S} satellites. 
Recently the {\it Fermi}-LAT has released a new measurement of the IGRB spectrum from 100 MeV to 820 GeV at Galactic latitude ($b$) $|b|>20^{\circ}$ \refcite{Ackermann:2014usa}. The LAT has also measured the Extragalactic $\gamma$-ray background (EGB) which is the sum of the IGRB and the flux from detected sources.
For the first time a deviation from the power-law shape in the high-energy part of the EGB and IGRB has been observed as an exponential cut off with a break energy of about $E=280$ GeV. 

The origin of the IGRB is not fully understood and a part of this residual is thought to arise from unresolved point sources.
Since Active Galactic Nuclei (AGNs), Star Forming Galaxies (SFGs) and pulsars are detected by the LAT these source populations are expect to explain the most of the IGRB (see Ref~\refcite{Fornasa:2015qua} for a review).
In Sec.~\ref{sec:interp} we will show that this hypothesis provides a viable explanation of the IGRB and EGB spectrum as found in Refs~\refcite{DiMauro:2013zfa,Ajello:2015mfa,DiMauro:2015tfa}. Moreover, I will show that the cut off is compatible with absorption of $\gamma$ rays from AGN due to their interaction with the Extragalactic Background Light (EBL).

The high sensitivity, the wide field of view together with observation of the sky in a large energy range enabled the {\it Fermi}-LAT to detect over 3000 sources during four years of operation \refcite{2015ApJS..218...23A}. 
A different strategy for the detection of $\gamma$ rays is followed by imaging air Cherenkov telescopes (IACTs). These ground-based experiments have a good angular resolution to detect and study $\gamma$-ray sources, but a limited field of view, and operate at $>$50--100\,GeV.

The {\it Fermi}-LAT Collaboration has recently released a new event-level analysis called Pass 8 that increases significantly the sensitivity of the LAT resulting in an acceptance improvement at least of 100$\%$ below 100 MeV and about 25$\%$ above 1 GeV \refcite{Atwood:2013dra}. These improvements are also quite significant at very high-energy (VHE) where at 50\,GeV and 500\,GeV the acceptance increases by a factor of 20$\%$ and 50$\%$ respectively.
Given the sensitivity of Pass 8 and with the goal of filling the gap with the IACTs, the {\it Fermi} Collaboration has prepared the second catalog of hard {\it Fermi}-LAT sources (2FHL) \refcite{Ajello:2015bda}.
The 2FHL catalog has been derived from 80\,months of Pass 8 data photons with reconstructed energy in the 50 GeV$-$2 TeV range. The $\gamma$-ray sky shines at this energies with approximately 61000 photons, which in turn led to the detection of 360 sources. 
The 2FHL catalog contains, at $|b|>10^{\circ}$, 70\% of sources associated to BL Lac population, 23\% are uncertain blazar type (BCU) and unassociated sources and only 7\% are associated to other source populations (4\% of them are FSRQs).
Due to their characteristics (frequency of the synchrotron peak and photon index) almost all BCU and unassociated sources are expected to be BL Lacs therefore the fraction of blazars in the 2FHL catalog is about 97\% (93\% BL Lacs and 4\% FSRQs).
In Sec.~\ref{sec:high} we are going to use this new catalog to derive the intrinsic source count distribution of blazars and demonstrate that the LAT is able to resolve almost all the EGB at $E>50$ GeV in terms of this source population.

\section{Interpreting IGRB and EGB data with source population studies}
\label{sec:interp}
Since the {\it Fermi}-LAT has measured thousands of $\gamma$-ray sources, a possible interpretation for the IGRB would be that it comes from the superposition of flux from unresolved sources. 
The {\it Fermi}-LAT catalogs are populated by AGNs, SFGs and pulsars and large efforts have been done by the scientific community to predict their contribution to the isotropic diffuse \refcite{Fornasa:2015qua}.

Blazars, AGN with the jet aligned with the line of sight, costitute the most numerous source population in {\it Fermi}-LAT catalogs (see e.g.~Refs~\refcite{2015ApJS..218...23A}) and they are therefore expected to give a large contribution to the IGRB (between 10\% and 40\% \refcite{DiMauro:2013zfa,2012ApJ...751..108A,Ajello:2013lka}).
According to the presence or absence of strong broad emission lines in their optical/UV spectrum, blazars are traditionally divided into flat-spectrum radio quasars (FSRQs) and BL Lacertae (BL Lacs) respectively \refcite{1995ApJ...444..567P}.
Blazars are also classified using the frequency of synchrotron-peak, into low-synchrotron-peaked (LSP), intermediate-synchrotron-peaked (ISP) or as high-synchrotron-peaked (HSP) blazars.
FSRQs are almost all LSP blazars and they only contribute to the low-energy part of the IGRB (between 8\% and 11\% below 10 GeV) \refcite{2012ApJ...751..108A}.
BL Lacs on the other hand are for one half LSP and ISP and the other half HSP sources and they contribute for about 10\% to the low energy part of the IGRB \refcite{DiMauro:2013zfa,Ajello:2013lka} while they explain almost entirely the IGRB at energy larger than 100 GeV \refcite{DiMauro:2013zfa}.

AGNs with jet misaligned with respect to the line of sight (MAGN) have weaker luminosities due to Doppler boosting effects \refcite{1995ApJ...444..567P} and only a dozen of MAGNs have been detected in {\it Fermi} catalogs. Taking into account a correlation between radio and $\gamma$-ray luminosities, the unresolved population of MAGN has been found to be very numerous and they are so expected to explain a large fraction of the IGRB (on average $30$-$40\%$) \refcite{2011ApJ...733...66I,DiMauro:2013xta}. However given the small catalog of $\gamma$-ray detected sources these predictions have large uncertainties and the MAGN unresolved emission can explain from 10 to 100\% of the IGRB.

SFGs are objects where the star formation rate is intense and $\gamma$ rays are produced from the interaction of cosmic rays with the interstellar medium or interstellar radiation field.
SFG $\gamma$-ray emission is very dim and the LAT has detected so far only nine individual galaxies \refcite{2012ApJ...755..164A}. 
Many more SFGs have been detected at infrared and radio wavelengths hence a numerous population of unresolved SFGs should exist and contribute to the IGRB ($4\%$ and $23\%$) as found in Ref.~\refcite{2012ApJ...755..164A} using correlations between the infrared or radio band with $\gamma$-ray emission.

Pulsars are the most promising Galactic contributors to the IGRB. In fact about 160 pulsars have been detected in the 3FGL catalog \refcite{2015ApJS..218...23A}. Pulsars are divided, according to the rotation period $P$, into young ($P>30$ ms) and millisecond pulsars (MSPs). Young pulsars are mostly concentrated along the Galactic plane so they can not give a contribution to the IGRB while the most of detected MSPs are at $|b|>10^{\circ}$.
However in a recent paper \refcite{Calore:2014oga} it has been estimated that unresolved MSPs can at most contribute with a few \% to the IGRB.

FSRQ, BL Lac blazars together with MAGN and SFG sources have been showed to give a possible explanation of the IGRB and EGB intensity and spectrum in the entire energy range.
This interpretation has been statistically tested for the first time in Refs~\refcite{Ajello:2015mfa,DiMauro:2015tfa} with a fit to IGRB and EGB data and including in the analysis also the theoretical uncertainties for the $\gamma$-ray emission of each population. In Ref~\refcite{DiMauro:2015tfa} it is also reported how the significance of this interpretation depends on the Galactic foreground model (GFM) used to derive the data \refcite{Ackermann:2014usa}. Although Refs.~\refcite{DiMauro:2013zfa,Ajello:2015mfa,DiMauro:2015tfa} consider different contribution for the blazar population, they both conclude that the EGB and IGRB {\it Fermi}-LAT spectra are statistically consistent with $\gamma$-ray emission from those extragalactic populations. 
An independent confirmation of this interpretation comes from Ref~\refcite{DiMauro:2014wha} where the angular power measured by the LAT \refcite{2012PhRvD..85h3007A} is explained by anisotropy produced by the same source populations.

\begin{figure}[h]
\begin{center}
{\includegraphics[width=2.4in]{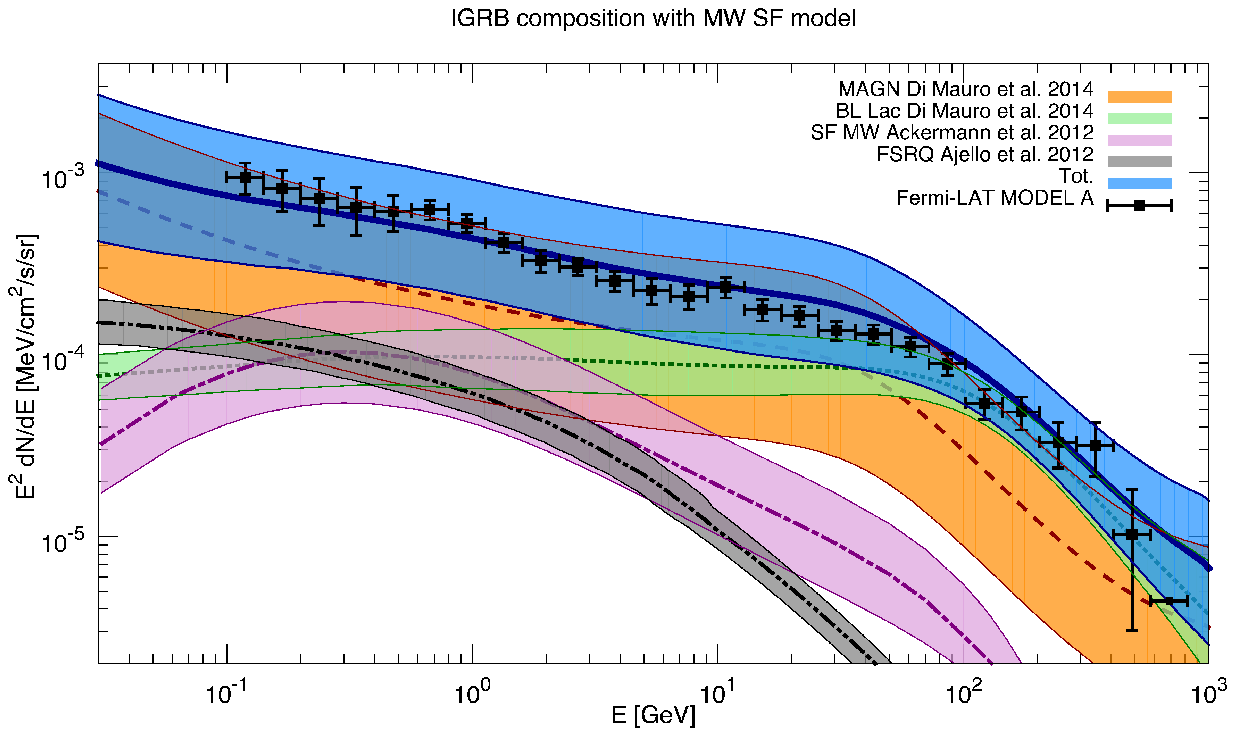}}
{\includegraphics[width=2.4in]{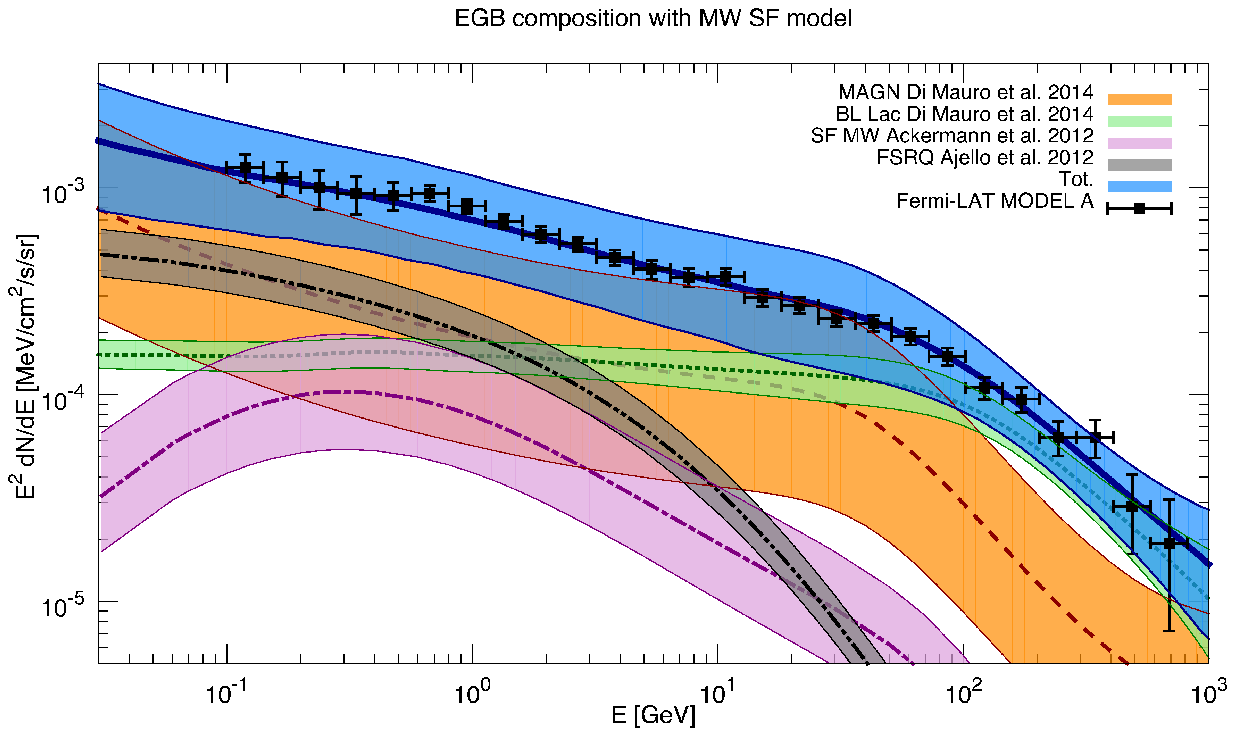}}
 \caption{Left (right) panel: average and uncertainty band for the $\gamma$-ray emission from unresolved (total=unresolved+resolved) MAGN, FSRQ, BL LAC and SFG sources, along with data for the IGRB (EGB).}
\label{fig:igrb_egb_theo}
\end{center}
\end{figure}

In Fig.~\ref{fig:igrb_egb_theo}, for example, we show the contribution of the above cited extragalactic populations (MAGN \refcite{DiMauro:2013xta}, FSRQ  \refcite{2012ApJ...751..108A}, BL Lac \refcite{DiMauro:2013xta} and SFG sources \refcite{2012ApJ...755..164A}) to the {\it Fermi}-LAT IGRB and EGB data derived with GFM A \refcite{Ackermann:2014usa}. We display the average and the $1\sigma$ error for each population and also the total contribution that, within the uncertainty band, is consistent with $\it Fermi$-LAT measurements.
FSRQ contribution is focused in the low part of data while BL Lac emission is more significative at $E>10$ GeV. This is due to the fact that FSRQs are at most LSP while BL Lacs are HSP blazars.
MAGN population, which is thought to be the parent misaligned population of blazars, contains LSP, ISP and HSP AGNs and they contribute in the entire energy range.
Finally SFG flux, because the $\gamma$-ray production is mostly due to $\pi^{0}$ decay, is peaked at 0.1-1 GeV.
The high-energy cut off, that for the first time the LAT instrument has measured, can be explained by the absorption of energetic $\gamma$ rays due to their interaction with the EBL. In this part of the IGRB in particular the data are entirely explained by BL Lac blazars \refcite{DiMauro:2013xta}.


\section{Resolving the extragalactic $\gamma$-ray background at $E>50$ GeV with {\it Fermi}-LAT}
\label{sec:high}
In this section we give an overview of the analysis performed by the {\it Fermi}-LAT Collaboration \refcite{TheFermi-LAT:2015ykq} where the intrinsic source count distribution of the 2FHL catalog blazars, at $E>50$ GeV and $|b|>10^{\circ}$, has been derived \refcite{TheFermi-LAT:2015ykq}.
Realistic simulations of the $\gamma$-ray sky have been performed considering the Galactic diffuse and isotropic emissions and the flux from an extragalactic population of point sources placed isotropically in the sky.
The photon index $\Gamma$ of simulated sources has been extracted from a Gaussian distribution with average value 3.2 and standard deviation 0.7 in order to reproduce the same spectral energy distribution observed in 2FHL catalog blazars.
The point source fluxes on the other hand have been drawn from a differential flux distribution $dN/dS$ given by a broken power-law where $\alpha_1$ and $\alpha_2$ are the slopes above and below the flux break $S_b$.
The slope of the flux distribution above the break flux is the one observed in the 2FHL, namely an euclidean distribution with $\alpha_1=5/2$. 
We have considered the same live time history and event selection used in the 2FHL catalog \refcite{Ackermann:2015uya} and we refer to Ref.~\refcite{TheFermi-LAT:2015ykq} for the full details of this analysis.

In order to derive the shape of $dN/dS$ below the flux threshold of the catalog, we have taken into account a photon fluctuation analysis that permitted us to probe the flux distribution to the level where sources contribute on average 0.5\,photons each.
This method employs the pixel count distribution, i.e. the number of pixel in function of the number of photons.
We simulated the point source population changing the flux break and the index below the break in the following range: $S_b\in[1\times10^{-12},2\times10^{-12}]$ ph cm$^{-2}$ s$^{-1}$ and $\alpha_2 \in [1.3,2.7]$.
For each realization of $S_b$ and $\alpha_2$ we derived 30 simulations and we binned the simulated and the real sky map with the HEALPix tool \refcite{2005ApJ...622..759G} taking a resolution of order 9 which translates into a pixel size of about $0.11^{\circ}$. We checked that using an order 8 our results did not change. 
Finally we compared for each couples of $S_b$ and $\alpha_2$ the average pixel count distribution of the simulations with the one of the real sky making a chi-square analysis.
The result of the fit is that the break flux is limited to the range between $S_b \in [8 \times 10^{-12},1.5 \times 10^{-11}]$\,ph cm$^{-2}$ s$^{-1}$ while the index below the break is in the range $\alpha_2\in[1.60,1.75]$.
The best configuration has a break flux at $1\times 10^{-11}$ ph cm$^{-2}$ s$^{-1}$ and a slope $\alpha_2=1.65$.

We employed the photon fluctuation analysis also to test a re-steepening of the source count distribution as it might occur if a source population of faint source, for instance SFGs, emerges in the flux distribution.
An additional flux break $S_{b,2}$ has been added with a fixed value of the slope $\alpha_3=2.50$ below $S_{b,2}$. We then moved the position of $S_{b,2}$ and we calculated the upper limit for this second break that worsens at $3\sigma$ the chi-square of the best-fit cited previously. This upper limit is $S_{\rm{lim}} \approx 7\times10^{-13}$ ph cm$^{-2}$ s$^{-1}$ and this faint population would exceed the EGB intensity at fluxes of $\sim 7\times10^{-14}$ ph cm$^{-2}$ s$^{-1}$.

\begin{figure}[h]
\begin{center}
{\includegraphics[width=2.4in]{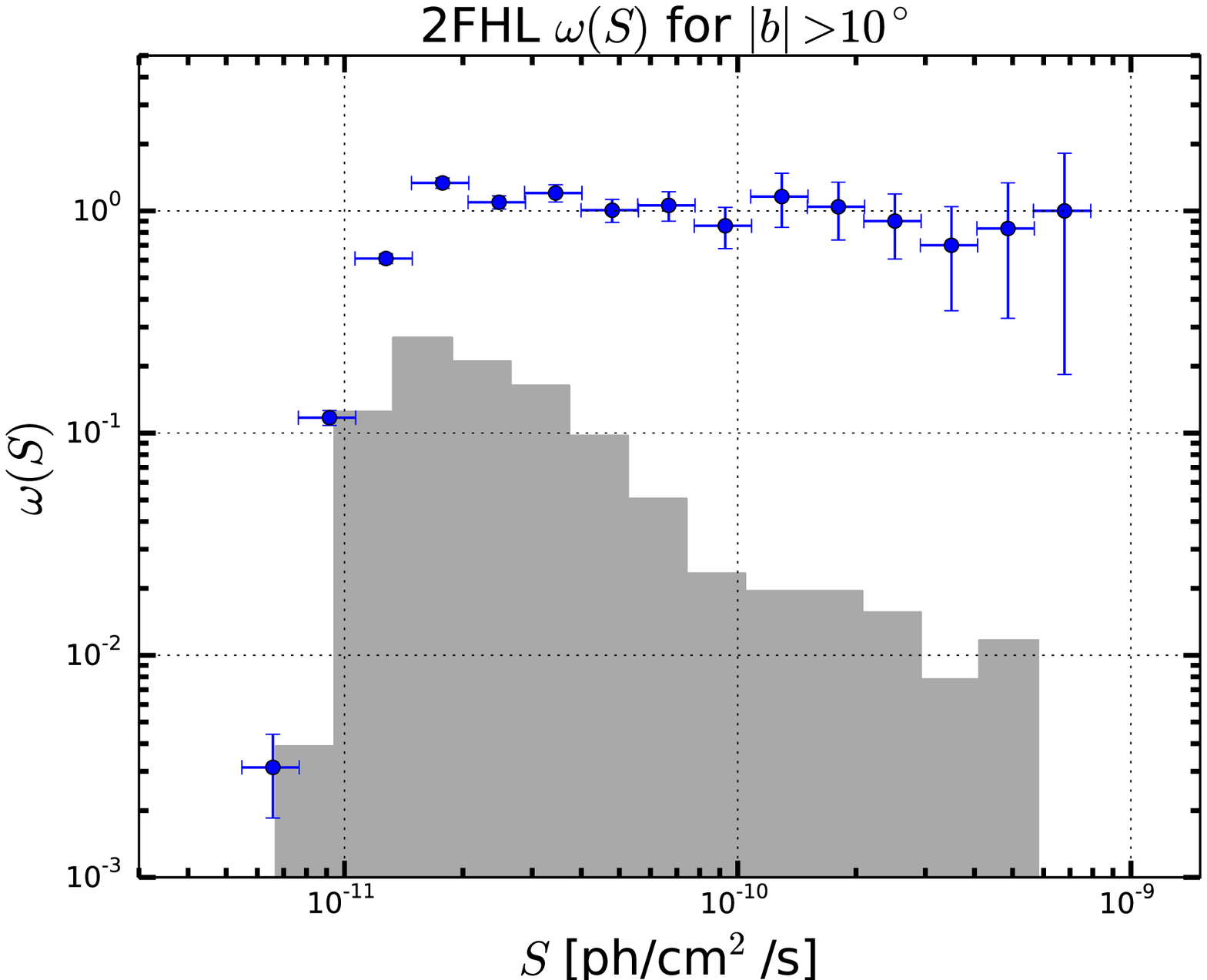}}
{\includegraphics[width=2.4in]{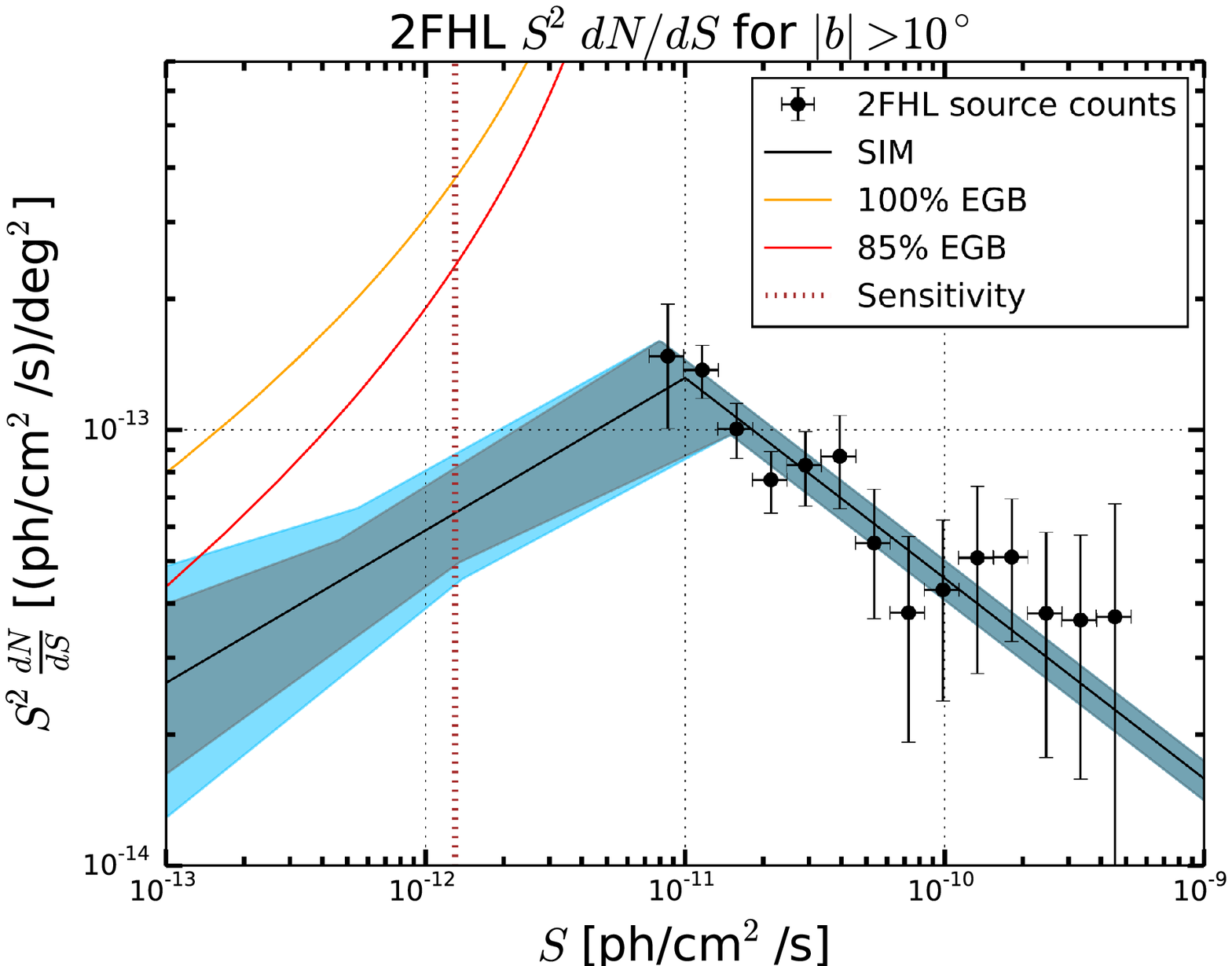}}
 \caption{Left Panel: detection efficiency $\omega(S)$ (blue points) as a function
of source flux and normalized distribution of source fluxes detected in 2FHL (grey shaded histogram).
Right Panel: intrinsic $S^2 dN/dS$ distribution of the 2FHL catalog (black points) together with the best-fit ($1\sigma$ and $3\sigma$) model with the black solid line (grey and cyan bands) as derived from the photon fluctuation analysis. The vertical brown dotted line is the sensitivity of the photon fluctuation analysis. and the orange and red curves indicate 85\% and 100\% level of the EGB intensity above 50 GeV.}
\label{fig:efficiency}
\end{center}
\end{figure}

Once we inferred the flux distribution of point sources, we have performed 10 simulations of the sky above 50 GeV and we analyzed them as done in the 2FHL catalog \refcite{Ackermann:2015uya}.
The number of detected sources at $|b| > 10^{\circ}$ is $271 \pm 18$ sources, in good agreement with the 253 sources in the 2FHL.
These results have been then used to derive the detection efficiency $\omega(S)$: the probability to detect a source with a flux $S$. The detection efficiency is derived for each flux bin as the ratio of detected sources and the number of simulated sources in that bin. The result is displayed in the left panel of Fig.~\ref{fig:efficiency} together with the histogram of the flux for the 2FHL catalog sources. The efficiency is equal to one for $S>2 \times 10^{-11}$ ph cm$^{-2}$ s$^{-1}$ meaning that the LAT detects any source for this flux range. Below this threshold $\omega(S)$ drops and at $\sim 1 \times 10^{-11}$ ph cm$^{-2}$ s$^{-1}$ the LAT misses 80$-$90\% of the detections.

The estimation of the detection efficiency gives the chance to correct the observed source count distribution and derive the intrinsic $dN/dS$ simply dividing the observed $dN/dS$ by $\omega(S)$.
In the right panel of Fig.~\ref{fig:efficiency} we show the average intrinsic source count distribution together with the $1$ and $3\sigma$ uncertainty bands derived in the photon fluctuation analysis. The intrinsic $dN/dS$ data of the 2FHL catalog are also reported together with the 85\% and 100\% levels of the EGB calculated extrapolating the flux distribution below the break with different values of $\alpha_2$.

The integration of the source count distribution $N(>S)=\int_S dN/dS' dS'$ gives the cumulative source count distribution $N(>S)$. Given the sensitivity of the photon fluctuation analysis ($\sim 1 \times 10^{-12}$ ph cm$-2$ s$^{-1}$) we are able to predict the number of sources below the threshold of the catalog and apply this exercise for instance to the CTA experiment.
For the predicted 5mCrab sensitivity reachable by CTA in 240 hours in the most sensitive pointing strategy \refcite{2013APh....43..317D} we predict $200 \pm 45$ sources detected in one quarter of the full sky.

We have also derived the contribution of point sources, mostly blazars, to the EGB integrating the source count distribution $I = \int^{\infty}_0 S' dN/dS' dS'$.  The total integrated flux from point sources is $2.07^{+0.40}_{-0.34}\times
10^{-9}$ ph cm$^{-2}$ s$^{-2}$ sr$^{-1}$ which constitutes $86^{+16}_{-14}\%$ of the EGB above 50GeV estimated in Ref.~\refcite{Ackermann:2014usa}. 
This result has important consequences because constrains all other source populations and diffuse mechanisms to contribute only with $14^{+14}_{-16}\%$ of the EGB at $>50$ GeV.

\section{Conclusions}
We summarized in this paper the key points for the interpretation of {\it Fermi}-LAT IGRB and EGB data \refcite{Ackermann:2014usa} in terms of $\gamma$-ray emission from FSRQ, BL Lac, MAGN and SFG sources.
This interpretation stands regardless the GFM employed for the derivation of IGRB and EGB datasets and the systematic uncertainties associated to emission from these extragalactic sources. 

Thanks to the large improvement of Pass 8 the {\it Fermi}-LAT has created a catalog of sources above 50 GeV \refcite{Ajello:2015bda}. We showed that, taking advantage of a photon fluctuation analysis, we derived the intrinsic source count distribution of blazars in this catalog up to almost one order of magnitude below the catalog sensitivity. Integrating the derived $dN/dS$, the $86^{+16}_{-14}\%$ of the EGB is resolved by blazars.

This implies a strong bound of $14^{+14}_{-16}\%$ for the contribution to the EGB from other source populations or diffuse processes, such as ultra high-energy cosmic rays (see e.g.~Ref.~\refcite{2011PhRvD..84h5019A}). Moreover, this limit would translate in severe constraints on the annihilation cross section or decay time of high-mass dark matter particles producing $\gamma$ rays  \refcite{DiMauro:2015tfa,Ajello:2015mfa,Calore:2013yia}. 
Finally, the $\gamma$-ray emission mechanism of SFGs, due to hadronic interaction of cosmic rays with the interstellar medium, produces also VHE neutrinos detectable by Ice Cuce. Therefore, the bound we have found for the $\gamma$-ray flux of SFGs translates in a strong evidence against the SFG interpretation of astrophysical neutrino flux measured by Ice Cube \refcite{Bechtol:2015uqb}.

{\footnotesize
\section*{Acknowledgments}
The \textit{Fermi}-LAT Collaboration acknowledges support for LAT development, operation and data analysis from NASA and DOE (United States), CEA/Irfu and IN2P3/CNRS (France), ASI and INFN (Italy), MEXT, KEK, and JAXA (Japan), and the K.A.~Wallenberg Foundation, the Swedish Research Council and the National Space Board (Sweden). Science analysis support in the operations phase from INAF (Italy) and CNES (France) is also gratefully acknowledged.
}

\bibliographystyle{ws-procs975x65}
\bibliography{ws-pro-sample}

\end{document}